\newcommand{\be}{\begin{equation}}
\newcommand{\ee}{\end{equation}}
\newcommand{\br}{\begin{eqnarray}}
\newcommand{\er}{\end{eqnarray}}

\documentstyle[aps,psfig,amssymb,multicol]{revtex}

\begin{document}

\title{On the P-representable subset
of all bipartite Gaussian separable states}
\author{Marcos C. de Oliveira\footnote{marcos@df.ufscar.br}}
\address{%\\
 Instituto de F\'\i sica ``Gleb Wataghin'',  Universidade Estadual de Campinas,\\
  13083-970, Campinas - SP, Brazil.}
\date{\today}

\maketitle
%\vspace{2cm}
%\begin{figure}
%\centerline{$\;$\hskip 0 truecm\epsfig{figure=uq.eps,height=2 cm}}
%\end{figure}
 \begin{abstract}
%{\it Key words}: entanglement; quantum evolution; non-classical states, quantum information.
P-representability is a necessary and sufficient condition for
separability of bipartite Gaussian states only for the special
subset of states whose covariance matrix are $Sp(2,R)\otimes
Sp(2,R)$ locally invariant. Although this special class of states
can be reached by a convenient $Sp(2,R)\otimes Sp(2,R)$
transformation over an arbitrary covariance matrix, it represents
a loss of generality, avoiding inference of many general aspects
of separability of bipartite Gaussian states.

\end{abstract}
\begin{multicols}{2}
\narrowtext \tighten

%%%%%%%%%%%%%%%%%%%%%%%%%%%%%%%%%%%%%%%%%%%%%%%%%%%%
%\section{huge\bf Problem Statement}}
%%%%%%%%%%%%%%%%%%%%%%%%
\section{Introduction}
%%%%%%%%%%%%%%%%%%%%%%%%

In the recent years the question to whether a given quantum state
is separable or entangled has become central to the quantum
information and to the quantum optics communities. Mostly because
fault-tolerant quantum information protocols, such as quantum
computation and quantum teleportation are completely dependent on
the ability to prepare pure (or close to pure) entangled states
\cite{nielsen}. Recent attention however has been centered on
continuous variable versions of quantum communication protocols,
such as the unconditional quantum teleportation
\cite{braukimble,furusawa}, whose efficiency rests on the ability
to generate entangled states of systems with infinite dimensional
Hilbert spaces. Bipartite systems with finite Hilbert space have
been exhaustively investigated in order to achieve a precise
quantification of entanglement. Peres \cite{peres} and Horodecki
\cite{horodecki} demonstrated that a necessary and sufficient
condition for separability of bipartite systems with Hilbert
spaces of dimension $\le 2\otimes 3$ is  the {\it positivity} of
the partial transpose of the system density matrix. On the other
hand, algebraic similarities between bipartite states with Hilbert
space
 of dimension $ 2\otimes 2$ and bipartite Gaussian states
 (described by $4\times 4$ covariance matrices) allows the extension of the positivity criterion
 to those special continuous variable states as firstly developed in Refs.
 \cite{simon,duan}, and considered afterwards in discussions on entanglement of
Gaussian bipartite states (e.g.
\cite{paulinax,dodonov,englert,fiurasek,esantos,daffer,kimmunro,paulina,adesso,plenio,marcos}).

Of particular importance is the connection between Glauber
P-representability of a bipartite quantum state and separability
\cite{englert}. A {\it P-representable state} is the one that is
represented by a positive Glauber P distribution function
$P(\alpha,\beta)$, which is less(or equally)-singular than the
delta distribution,  such as \be \label{prep}\rho=\int d\alpha^2
d\beta^2
P(\alpha,\beta)|\alpha,\beta\rangle\langle\alpha,\beta|.\ee Under
this condition $P(\alpha,\beta)$ assumes the structure of a
legitimate probability distribution function over an ensemble of
states, allowing the connection between separability and
classicality. However, although any P-representable bipartite
state is separable, as can be immediately seen by the P
representation definition, the inverse is not necessarily true.
P-representability and separability are completely equivalent only
for Gaussian states with locally $Sp(2,R)\otimes Sp(2,R)$
invariant covariance matrices. Since any covariance matrix can be
brought to this invariant form under appropriate $Sp(2,R)\otimes
Sp(2,R)$ transform, P-representability and separability have been
misleadingly accepted as one-to-one equivalent properties of
bipartite Gaussian states. The purpose of the present paper is to
give a complete classification of the set of all bipartite
Gaussian separable states (BGSS). Particularly we show that
P-representable Gaussian bipartite states form a subset of BGSS
with locally $Sp(2,R)\otimes Sp(2,R)$ invariant form. In Section
II we begin by a revision of some necessary properties of
bipartite Gaussian states and in Sec. III we give the necessary
and sufficient conditions for the state to be separable. In Sec.
IV we discuss the P-representability of those states and show that
they actually form a subset of the separable states. In Sec. V we
describe the the unitary $Sp(2,R)\otimes Sp(2,R)$ map that
connects the two sets and finally in Sec. VI a conclusion encloses
the paper.

\section{Bipartite Gaussian States}

Any bipartite quantum state $\rho$ is Gaussian (see e.g.
\cite{plenio,englert2}) if its symmetric characteristic function
is given by \be
C({\bbox{\eta}})=Tr[D({\bbox{\eta}})\rho]=e^{-\frac12{\bbox{\eta}^\dagger}{\bf
V}{\bbox{\eta}} }, \ee where
$D(\bbox{\eta})=e^{-\bbox{\eta}^\dagger{\bf E}{\bf v}}$ is a
displacement operator in the parameter four-vector
$\bbox{\eta}$-space, with \be\bbox{\eta}^\dagger=\left(\eta_1^*,
\eta_1, \eta_2^*, \eta_2\right),\;\; {\bf v}^\dagger=
\left(a_1^\dagger, a_1, a_2^\dagger, a_2\right),\ee
%\be{\bbox{\eta}}= \left(\begin{array}{c}\eta_1\\
%\eta_1^*\\ \eta_2\\\eta_2^*\end{array}\right),\;\;{\bf v}= \left(\begin{array}{c}a_1\\
%a_1^\dagger\\ a_2\\a_2^\dagger\end{array}\right), \ee
 and \be {\bf
E}=\left(\begin{array}{c c}{\bbox{Z}}&{\bf 0}\\{\bf 0}&
{\bbox{Z}}\end{array}\right),\;\;\;
{\bbox{Z}}=\left(\begin{array}{c c}{1}&{0}\\{0}&
{-1}\end{array}\right),\ee where $a_1$ ($a_1^\dagger$) and $a_2$
($a_2^\dagger$) are annihilation (creation) operators for party 1
and 2, respectively. ${\bf V}$ is the Hermitian  $4\times 4$
covariance matrix with elements
$V_{ij}=(-1)^{i+j}\langle\{v_i,v_j^\dagger\}\rangle/2$, which can
be decomposed in four block $2\times 2$ matrices, \be\label{var}
{\bf V}=\left(\begin{array}{c c}{\bf V_1}&{\bf C}\\{\bf
C}^\dagger& {\bf V_2}\end{array}\right),\ee where ${\bf V_1}$ and
${\bf V_2}$ are Hermitian matrices containing only local elements
while ${\bf C}$ is the correlation between the two parties. Any
covariance matrix must be positive semidefinite (${\bf V}\ge 0$),
furthermore the generalized uncertainty principle\be
\label{cond}{\bf V}+\frac 12 {\bf E}\ge 0,\ee  must also be
applied. Those general positivity criteria can be decomposed into
block using matrix positivity properties. There are many ways to
check a matrix positivity, such as the positivity of the matrix
determinant. However this is not a necessary condition. A reliable
and convenient way to check the covariance matrix positivity is
through the following block Schur decomposition \cite{horn}: {\it
Any Hermitian matrix is positive if and only if any principal
block matrix is also positive, or, if its upper left block and the
block's Schur complement are also positive}. So that for the
covariance matrix (\ref{var}), ${\bf V}\ge 0$ only if \be{\bf
V_1}\ge 0,\ee
 and the Schur complement of ${\bf
V_1}$ \be S({\bf V_1})\equiv{\bf V_2}-{\bf C}^\dagger({\bf
V_1})^{-1}{\bf C}\ge 0.\ee It is interesting to observe that Schur
complements of block matrices representing Gaussian states
covariances, such as above, embodies a manifestation of a physical
operation when considering partial projections onto Gaussian
states \cite{eisert,jaromir}.

 Through the Schur decomposition the physical positivity
criterion (\ref{cond}) applies only if \be \label{xcond}{\bf
V_1}+\frac 12 {\bf Z}\ge 0,\ee and
 \be \label{xcond2}\left({\bf V_2+\frac 12 {\bf Z}}\right)-{\bf C}^\dagger\left({\bf
V_1}+\frac 12 {\bf Z}\right)^{-1}{\bf C}\ge 0.\ee By explicitly
writing
  \br {\bf V}=\left(\begin{array}{c c
c c}n_1&m_1&m_s&m_c\\m^*_1& n_1&m^*_c&m^*_s\\
%-&-&-&-&-\\
m^*_s&m_c&n_2&m_2\\ m_c^*&m_s&m_2^*&n_2\end{array}\right) ,\er we
can further
simplify the generalized uncertainty (\ref{xcond}) and (\ref{xcond2}) to \br \label{sep}n_1&\ge& \sqrt{|m_1|^2+\frac 1 4},\\
\mbox{and}\nonumber\\\label{2sep} n_2&\ge&
\frac{s}{d}+\sqrt{\frac1 4
\left[\frac{\left||m_c|^2-|m_s|^2\right|}{d}-1\right]^2+|m_2-c|^2},\er
respectively, with\br
s&=&n_1\left(|m_c|^2+|m_s|^2\right)-m_cm_sm_1^*-m_c^*m_s^*m_1,\\
c&=&2n_1m_s^*m_c-m_c^2m_1^*-(m_s^*)^2m_1,\\
 d&=&n_1^2-\frac1 4-|m_1|^2.\er
 %In Eq. (\ref{2sep}) the signal $+$
 %($-$) applies for $|m_c|^2\ge|m_s|^2$ ($|m_c|^2<|m_s|^2$).

\section{Separability Bounds}

By mapping the positivity necessary and sufficient condition for
  dimension $2\otimes 2$ to bipartite systems of infinite dimension, Simon
\cite{simon} has discovered an elegant geometrical interpretation
of separability in terms of the Wigner distribution function for
the density operator. The Peres-Horodecki separability criterion
in the Simon framework reads: {\it if a bipartite density operator
is separable, then its Wigner distribution necessarily goes over
into a Wigner distribution under a phase space mirror reflection.}
The separability criterion can be understood as a valid
Wigner-class-conservative quantum map under local time reversal.
 Following \cite{simon} a necessary and sufficient
condition for a Gaussian quantum state to be separable, \be
\rho=\sum_k p_k \rho_k^A\otimes\rho_k^B \label{sepdef}\ee is that
its covariance matrix must satisfy\be {\bf \widetilde V}+\frac 12
{\bf E}\ge 0,\label{cond2}\ee under a partial phase space mirror
reflection (partial Hermitian conjugation) ${\bf\widetilde V}={\bf TVT}:$\be{\bf T v}={\bf v}_T=\left(\begin{array}{c}a_1\\
a_1^\dagger\\ a_2^\dagger\\a_2\end{array}\right),\ee with \be {\bf
T}=\left(\begin{array}{c c}{\bf I}&{\bf 0}\\{\bf 0}& {\bf
X}\end{array}\right),\;\; {\bf X}=\left(\begin{array}{c c}{ 0}&{
1}\\{1}& {0}\end{array}\right),\ee otherwise the state is
entangled. Similarly to the generalized uncertainty (\ref{cond})
decomposition,
 the separability condition (\ref{cond2}) is satisfied if and only
 if (\ref{xcond}) and
 \be \left({\bf XV_2X+\frac 12 {\bf Z}}\right)-{\bf X C}^\dagger\left({\bf
V_1}+\frac 12 {\bf Z}\right)^{-1}{\bf C X}\ge 0,\ee are both
satisfied, which explicitly implies in (\ref{sep}) and
 \br \label{sep2}n_2&\ge& \frac{s}{d}+\sqrt{\frac1 4
\left[\frac{\left||m_c|^2-|m_s|^2\right|}{d}+1\right]^2+|m_2-c|^2},\er
respectively.
% with the same signal convention of Eq. (\ref{2sep}).
We call the set of states $\rho$ that fall inside (\ref{sep}) and
(\ref{sep2}) the set BGSS of all bipartite Gaussian separable
states $\mathbb{S}$. Any state that does not follow those
inequalities is entangled, being it pure or not. Remark also that
purity is only reached when the equalities in (\ref{sep}) and
(\ref{2sep}) hold.

\section{P-representability of Gaussian states}
The very definition of a separable state (\ref{sepdef}) can be
written in a coherent state representation through the Glauber
P-function (\ref{prep}), but it is not obvious that
$P(\alpha,\beta)$ is a legitimate probability distribution
function. That is only reached if the state is P-representable,
i.e., if the P-function is non-negative and less (or equal)
singular than the delta distribution. In terms of the covariance
matrix, a quantum state is P-representable \cite{englert2} if
 \be {\bf V}-\frac 12 {\bf I}\ge
 0,\ee
 which in terms of
the upper left block matrix and its Schur complement writes as
 \be {\bf V_1}-\frac 12 {\bf I}\ge0,\ee and
 \be\left({\bf V_2-\frac 12 {\bf I}}\right)-{\bf C}^\dagger\left({\bf
V_1}-\frac 12 {\bf I}\right)^{-1}{\bf C}\ge 0.\ee Thus
\be\label{p} n_1\ge|m_1|+\frac12,\ee and \be \label{p2}n_2\ge
\frac{s^\prime}{d^\prime}+\frac{|m_2-c^\prime|}{d^\prime}+\frac12,\ee
with\br
s^\prime&=&(n_1-\frac12)\left(|m_c|^2+|m_s|^2\right)-m_cm_sm_1^*\nonumber\\
&&-m_c^*m_s^*m_1,\\
c^\prime&=&2(n_1-\frac 1 2)m_s^*m_c-m_c^2m_1^*-(m_s^*)^2m_1,\\
 d^\prime&=&(n_1-\frac1 2)^2-|m_1|^2.\er States that follow (\ref{p}) and (\ref{p2})
 form the set of all
 bipartite P-representable Gaussian states $\mathbb{P}$.
% Remark that the two conditions lead to different inequalities.

From an operator formalism for the density matrix, Englert and
W\'odkiewicz \cite{englert} have recently stated that
P-representability is equivalent to the separability condition,
for the specific symmetric situation where $m_1=m_2=m_s=0$,
$n_1=n_2=n$, and $m_c=m$, which indeed set
$\mathbb{P\rightleftharpoons S}$ as we see bellow. The generality
of their statement is justified only if $Sp(2,R)\otimes Sp(2,R)$
local operations are used to bring  those parameters to the
special symmetric class described above
  (see also \cite{englert2}). However, this particular situation does not represents a total
    equivalence
  between  $\mathbb{S}$ and the set of all P-representable states.
  In general the P-representability conditions,
   (\ref{p}) and (\ref{p2}), are more
  restrictive than the separability ones, (\ref{sep}) and (\ref{sep2}), respectively, as we now
  investigate.
%\begin{figure}
%\vspace{-0.20cm}
% \centerline{$\;$\hskip 0truecm\psfig{figure=fig1.ps,height=12cm}}\vspace{-6cm}
%\parbox{8cm}{\small Fig.1.
%Typical separability ({\bf S}) and P-representability ({\bf P})
%boundaries for $m_1=0$ and $m_2=1$.}
%\end{figure}

Firstly observe that (\ref{sep}) is less restrictive than
(\ref{p}), equaling only for $|m_1|=0$ or
$|m_1|\rightarrow\infty$, being enough to check if (\ref{p2})
dominates over (\ref{sep2}) for the simplest $|m_1|=0$ situation.
For that we make use of the knowledge that (\ref{2sep}) is always
stronger than (\ref{p2}), including the situation where $d=0$,
i.e., $n_1=1/2$. In such a case, the comparisons of the (\ref{p2})
lower bound to (\ref{sep2}) and to (\ref{2sep}) are equivalent and
thus if  (\ref{sep2}) is violated so is (\ref{2sep}). These
inequalities must satisfy
\be\label{ineq}(|m_2|+|m_c|^2)(|m_2|+|m_s|^2)\ge 0,\ee and since
the quantities involved  are always strictly positive the
criterion (\ref{ineq}) is always satisfied. The equality however
occurs only if $|m_2|=|m_c|^2=0$ or $|m_2|=|m_s|^2=0$, which then
set the equivalence $\mathbb{P\rightleftharpoons S}$ for the two following special $Sp(2,R)\otimes Sp(2,R)$ invariant forms for ${\bf V}$:\\
 {\it Invariant form 1:}
 \be
\left(\begin{array}{c c
c c}n_1&0&0&m_c\\0& n_1&m_c^*&0\\
%-&-&-&-&-\\
0&m_c&n_2&0\\ m_c^*&0&0&n_2\end{array}\right),\ee\\ {\it Invariant
form 2:}
 \be \left(\begin{array}{c c
c c}n_1&0&m_s&0\\0& n_1&0&m^*_s\\
%-&-&-&-&-\\
m^*_s&0&n_2&0\\ 0&m_s&0&n_2\end{array}\right).\ee Special forms 1
and 2 are locally $Sp(2,R)\otimes Sp(2,R)$ invariant covariance
matrices that form the $\mathbb{P\rightleftharpoons S}$ subset.
The separability and thus P-representability criterion is then
reduced to \be \left(n_1-\frac 12\right)\left(n_2-\frac
12\right)\ge|m_c|^2,\ee for the special form 1, and to \be
\left(n_1-\frac 12\right)\left(n_2-\frac 12\right)\ge|m_s|^2,\ee
for the special form 2. The physical condition of existence of a
general bipartite Gaussian state of the form 1 or 2 writes \be
\left(n_1-\frac 12\right)\left(n_2+\frac 12\right)\ge|m_c|^2,\ee
or \be \left(n_1-\frac 12\right)\left(n_2+\frac
12\right)\ge|m_s|^2,\ee respectively.
\begin{figure}\vspace{0.20cm}
 \centerline{$\;$\hskip 0truecm\psfig{figure=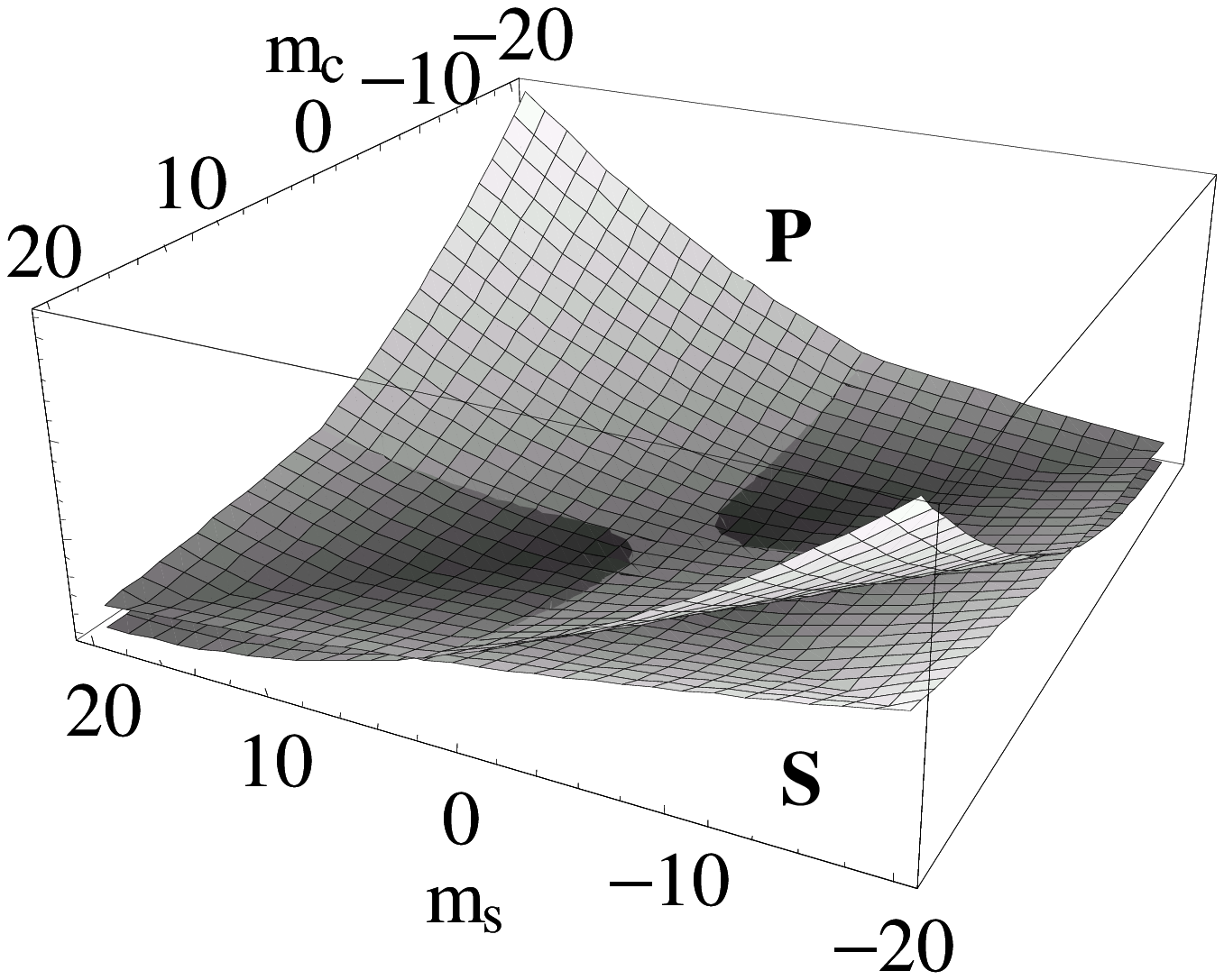,height=10cm}}\vspace{-5.5cm}
\parbox{8cm}{\small Fig.1. Typical Separability ({\bf S}) and P-representability ({\bf P}) boundaries
for $m_1=0.5$ and $m_2=1$. The shaded area where the {\bf P}-fold
is lower than the {\bf S}-fold does not represent physical quantum
states.}
\end{figure}

 {\it Remark 1:} There are
P-representable Gaussian operators that violate (\ref{2sep}),
which however do not represent any valid positive definite quantum
state. As an example in Fig. 1 through comparison of the limiting
bounds (\ref{sep2}) and (\ref{p2}), assuming all real coefficients
and setting $m_1=0.5$, and $m_2=1$. Only those states that lay in
or above the separable class are valid P-representable separable
states.

{\it Remark 2:} The special symmetric situation depicted in
Ref.\cite{englert,daffer} for the
 two-mode thermal squeezed state, where $m_1=m_2=m_s=0$, $n_1=n_2=n$, and
 $m_c=m$, is a particular example of the specific form 1, and thus a separable
 state in this case is always P-representable.

\section{$\mathbb{S}\rightleftarrows\mathbb{P}$ Mapping}

Any general covariance matrix can be mapped into one of those
invariant forms under appropriate $Sp(2,R)\otimes Sp(2,R)$
transform. In other words, it is possible to map $\mathbb{S}$ into
$\mathbb{P}$ such that \be \rho_{Sp}=U_L\rho_GU^{-1}_L,\ee be the
state obtained by the local unitary transform $U_L=U_1\otimes U_2$
over a general bipartite Gaussian density operator $\rho_G$
assuming \be U_L{\bf v}U^{-1}_L={\bf S_L} {\bf v},\;\;\; {\bf
S_L}=\left(\begin{array}{c c}{\bf S_1}&{\bf 0}\\{\bf 0}& {\bf
S_2}\end{array}\right),\ee with the condition ${\bf S_L}^{-1}={\bf
E S_L}^\dagger{\bf E}$. The new symmetric characteristic function
writes \br
C_{Sp}({\bbox{\eta}})&=&Tr[D({\bbox{\eta}})\rho_{Sp}]=Tr[U_L^{-1}
D({\bbox{\eta}})U_L\rho_{G}]\nonumber\\
&=&e^{-\frac12{\bbox{\eta}^{\dagger}}{\bf V}_{Sp}{\bbox{\eta}}
},\er with \be {\bf V}_{Sp}={\bf S_L}^{\dagger}{\bf V}{\bf
S_L}.\label{trans}\ee The transformed covariance matrix writes as
(\ref{var}), but with new block elements \br {\bf V_i^\prime}&=&
{\bf S_i}^{\dagger}{\bf V_i}{\bf S_i},\;\;
 {\bf C^\prime}= {\bf S_1}^{\dagger}{\bf C}{\bf S_2}.\er

 Assuming a local $Sp(2,R)$ transform as \br
 {\bf S_i}&\equiv&\left(\begin{array}{cc} e^{i\phi_i}\cosh\theta_i&
  e^{i\varphi_i}\sinh\theta_i\\e^{-i\varphi_i}\sinh\theta_i&e^{-i\phi_i}\cosh\theta_i
  \end{array}\right),\label{trans2}
\end{eqnarray}
 the condition to bring ${\bf V}$ to the invariant form 1:
 \be
\left(\begin{array}{c c
c c}\nu_1&0&0&\mu_c\\0& \nu_1&\mu_c^*&0\\
%-&-&-&-&-\\
0&\mu_c&\nu_2&0\\ \mu_c^*&0&0&\nu_2\end{array}\right),\ee is
obtained by setting $\phi_i+\varphi_i=-\mu_i+\pi$ and $\tanh
2\theta_i=|m_i|/n_i=|m_c|/|m_s|$, for $i=1,2$, respectively, where
$e^{-i\mu_i}=m_i/|m_i|$, and assuming (for $|m_c|\ge|m_s|$).

Now the condition to bring ${\bf V}$ to the invariant form 2:
 \be \left(\begin{array}{c c
c c}\nu_1&0&\mu_s&0\\0& \nu_1&0&\mu^*_s\\
%-&-&-&-&-\\
\mu^*_s&0&\nu_2&0\\ 0&\mu_s&0&\nu_2\end{array}\right),\ee
 is
immediately attained if $\phi_i+\varphi_i=-\mu_i+\pi$ also, but
now with $\tanh 2\theta_i=|m_i|/n_i=|m_s|/|m_c|$ (assuming
$|m_c|\le|m_s|$). Since both ${\bf V_1}^\prime$ and ${\bf
V_1}^\prime$ are proportional to the identity, they do not change
under unitary local rotations and the two invariant forms are then
connected through those operations. As such, the last two
conditions on $|m_c|$ and $|m_s|$ can be waved by appropriate
rotations.

The new transformed elements are \br
\nu_i&^=&\sqrt{n_i^2-|m_i|^2},\\
\mu_s&=&
e^{-i(\phi_1-\phi_2)}\frac{m_s}{|m_s|}\sqrt{|m_s|^2-|m_c|^2},\er
(for $|m_c|\le|m_s|$), and \be\mu_c=
e^{-i(\phi_1+\phi_2)}\frac{m_c}{|m_c|}\sqrt{|m_c|^2-|m_s|^2},\ee
(for $|m_c|\ge|m_s|$), which then turn explicit the four
invariants of the $Sp(2,R)\otimes Sp(2,R)$ group: $I_1=\det{\bf
V_1^\prime}$, $I_2=\det{\bf V_2^\prime}$,  $I_3=\det{\bf
C^\prime}$, and $I_4=Tr\left[{\bf V_1^\prime \bbox{Z} C^\prime
\bbox{Z} V_2^\prime \bbox{Z} (C^\prime)^\dagger \bbox{Z}}\right]$.

The general  $Sp(2,R)\otimes Sp(2,R)$ transformation (\ref{trans})
of the (\ref{trans2}) form is reached through the squeezing
operation $U_L=U_1 \otimes U_2:$ \be U_i=e^{i\frac{\hbar t} 2
\left(\kappa_i
{a_i^\dagger}^2e^{i\varphi_i}-\kappa_i^*{a_i}^2e^{-i\varphi_i}\right)},\ee
over the bipartite Gaussian state $\rho_{G}$, with $|\kappa_i|
t=\theta_i\equiv 2r_i$, the squeezing parameter associated with
the transformation on the mode $i$ and
$e^{i\phi_i}=\kappa_i/|\kappa_i|$. An important result is that
while all the BGSS set can be mapped into the P-representable set
by suitable $Sp(2,R)\otimes Sp(2,R)$ transforms, it is not
possible to restitute the original matrices ${\bf V_1}$ and ${\bf
V_2}$ with unitary rotations. That is only reached applying over
the squeezing operation. This is immediate from the two invariant
forms. Since both covariances reduced matrices ${\bf V_1^\prime}$
and ${\bf V_2^\prime}$ are proportional to the identity, unitary
rotations transform the invariant forms among themselves.

{\it Remark 3:} Through the $\mathbb{S}\rightleftarrows\mathbb{P}$
mapping, we have reached the special subset of locally
$Sp(2,R)\otimes Sp(2,R)$ invariant forms. However the general
separability condition can possibly be set equivalent to the
special P-representable subset under an appropriate nonlocal
operation forming the mapping $\mathbb{S}\rightarrow\mathbb{P}$.
Let us consider again the condition for separability
(\ref{cond2}). It can be equivalently written as\be {\bf V}+\frac
12 {\bf T E T}\ge 0.\label{cond2n}\ee Now let $U_{NL}$ be a
nonlocal operation:
\begin{eqnarray}
%\rho_{out}&=&B\rho_{in}B^\dagger,\label{1}\\
U_{NL}{\bf v}U_{NL}^\dagger &=&{\bf M} {\bf v},\label{2}
%{\bf M}&=&\left(\begin{array}{c c}{\bf R}&{\bf S}\\{\bf -S}^*&
%{\bf R}^*\end{array}\right)\\
% {\bf R}=\cos\theta\left(\begin{array}{cc} e^{i\phi_0}&0\\0&
%e^{-i\phi_0}\end{array}\right),&{\bf
%S}&=\sin\theta\left(\begin{array}{cc}e^{i\phi_1}&0\\0&e^{-i\phi_1}\end{array}\right).
%\\B|00\rangle&=&|00\rangle,\label{32}
\end{eqnarray}
where $\mathbf{M}$ is a general transformation matrix:
$\mathbf{M}\in$ Sp(4,R). Such a general $\mathbf{M}$, when acting
on (\ref{cond2n}) must leave {\bf V} invariant in form (${\bf
V^\prime}$), while ${\bf M^\dagger T E TM}$ must go necessarily to
$-\mathbf{I}$, such that (\ref{cond2n}) writes as \be {\bf
V^\prime}-\frac 12 {\bf I}\ge 0,\ee which is the transformed
P-representable subset condition. The Stone - von Neumann theorem
provides that if ${\bf M}$ exists it must be unitarily
implementable \cite{arvind}. Nonetheless, finding the
corresponding $U_{NL}$ operator may not be a simple exercise and
we leave this point for future research.

\section{Conclusion}

 In conclusion, we have derived a complete description of
  bipartite Gaussian separable states, and have proved that
 P-representable states form a subset of the set of all bipartite
 Gaussian separable states, existent only under special symmetry of
  the covariance matrix.
We can  state that for positive definite bipartite Gaussian
operators, which describe physical quantum states,
     {\it P-representability is a necessary and sufficient
condition for separability only for the subset of locally
$Sp(2,R)\otimes Sp(2,R)$ invariant Gaussian states}\cite{nota}. In
General $\mathbb{P} \subset\mathbb{S}$.

\acknowledgments{I am grateful to P. Marian and E. Santos to bring
Refs. \cite{paulinax} and \cite{esantos}, respectively, to my
knowledge. I want to thank A.Z. Khoury, K. Dechoum, M.K. Olsen, D.
Jonathan, P.H. Souto Ribeiro and V.V. Dodonov for many interesting
discussions. This work is partially supported by FAPESP and
Instituto do Mil\^enio de Informa\c c\~ao Qu\^antica (CNPq).}
%%%%%%%%%%%%%%%%%%%%%%%%%%%%%%%%%%%%%%%%%%%%%%%%%%%%%%%%%%%%%%%%%%

\end{multicols}

\begin{references}
\bibitem{nota} It is interesting to note that
the author of Ref.\cite{esantos} have arrived to similar
conclusions from a somewhat different approach on classicality and
Q-representation.
\end{references}
\end{document}